\documentclass{iopart}[12pt]
\usepackage{iopams}
\usepackage{epsfig}
\usepackage{times}
\usepackage{pstricks}
\usepackage{amsfonts,amsthm,eucal,amssymb}
\newcommand{\A}{\mbox{A}}
\newcommand{\W}{\mbox{W}}
\newcommand{\as}{\mathcal{A}}
\newcommand{\ws}{\mathcal{W}\,}

\begin{document}

\jl{1}

\title[Shape-invariant potentials with position-dependent mass through second order supersymmetry]{Shape-invariant quantum Hamiltonian with position-dependent effective mass  through second order supersymmetry}

\author{A~Ganguly$^{1,2}$\footnote{e-mail: gangulyasish@rediffmail.com}
        and L~M~Nieto$^1$\footnote{e-mail: luismi@metodos.fam.cie.uva.es}}

\address{$^1$Departamento de F\'{\i}sica Te\'orica, At\'omica y \'Optica, Univ. de Valladolid, 47071 Valladolid, Spain}

\address{$^2$City College (C.\ C.\ C.\ B.\ A.), University of Calcutta, 13 Surya Sen Street, Kolkata--700012, India}

\date{January 20, 2007}

\begin{abstract}
 Second order supersymmetric approach is taken to the system describing motion of a quantum particle
 in a potential endowed with position-dependent effective mass. It is shown that the intertwining relations between second order partner
 Hamiltonians may be exploited to obtain a simple shape-invariant condition.  Indeed a novel relation between potential and mass
 functions is derived, which leads to a class of exactly solvable model. As an illustration of our procedure, two examples
 are given for which one obtains whole spectra
 algebraically. Both shape-invariant potentials exhibit harmonic-oscillator-like or singular-oscillator-like
 spectra depending on the values of the shape-invariant parameter.
\end{abstract}



\section{Introduction}

The effective mass (EM) Schr\"odinger equation with position
dependent mass is a very useful model in many applied branches of
modern physics, e.~g.\  semiconductors \cite{basbk}, quantum dots
\cite{ser} , and $^3$He clusters \cite{bar}. In these cases, the
envelope wave function actually provides a macroscopic description
of the motion of carrier electrons with position dependent (or
equivalently material-composition dependent) mass. Recent interest
in this field [4--17] stems from extraordinary development of
nanostructure technology. This sophisticated technology of
semiconductor growth like molecular beam epitaxy technique, makes
production of ultrathin (nonuniform) semiconductor specimen a
reality nowadays. Consequently the study of such EM equation
becomes relevant for deeper understanding on the non-trivial
quantum effects observed in those nanostructures. Another area,
where this equation has extensive use, is the study of quantum
many-body systems in solids. For instance, the computational
limitation of the Green's function quantum Monte Carlo method is
removed by using the so-called pseudopotentials. The EM equation
appears in the process when one attempts to replace the nonlocal
(and thus disadvantageous) angular-momentum projection operator of
the pseudopotentials \cite{bac}.

On the other hand, supersymmetric (SUSY) approach \cite{susybook}
has been proved as a  powerful tool in quantum mechanics
identifying the isospectral partners in bosonic and fermionic
sectors, and thereby generating a hiererchy of solvable
Hamiltonians \cite{as1}. In the standard SUSY approach for EM
Hamiltonians the ladder operators are taken as first order
derivative operators similar to constant mass (CM) case, but now
they depend on both superpotential and mass function.  As a result
one obtains two partner EM potentials with the same effective mass
sharing identical spectra up to the zero mode of supercharge. In
the literature a  generalization of standard SUSY known as higher
derivative SUSY (HSUSY) or $\mathcal{N}$-fold SUSY, find a good
usage for CM models [21--31]. This method keeps the basic
superalgebra and differs from standard first order SUSY in that
the supercharges are represented as $\mathcal{N}$-th order
($\mathcal{N}>1$) differential operators. Recently this higher
order approach has been extended for EM Hamiltonian \cite{tan1}.
When one is interested in the solvability of the Hamiltonian,
shape-invariance (SI) is an important criteria \cite{gen}, because
in this case one can obtain full spectra by successive application
of raising operator. For instance, some classes of SI Hamiltonian
have been discovered recently through first order SUSY formalism
\cite{plas,dut2,bag}.

 Our purpose in the present work is to extend the concept of SI to
 second order SUSY (SSUSY) in the effective mass framework. This type of extension has been done
 very recently \cite{and2} for CM case. We
 have derived a new relation between the potential and mass functions, which gives an exactly-solvable (ES) model
 satisfying a simple SI condition.
 Two specific examples are given, one is hyperbolic
 and other is algebraic. Both these Hamiltonians are new ES
 models, because their wave functions and energy eigenvalues
 can be obtained by purely algebraic means. For certain values of the characterizing parameter these potentials
 acquire an inverse-square singularity at the center. Such type of singular potentials
 has important applications in various fields e.~g.\ molecular and
 high energy nuclear physics \cite{wil}. In the
 constant mass limit both of these reduce to well-known harmonic oscillator or singular
 oscillator depending on the values of the SI parameter.

 In Section~\ref{firstorder} we set up first order SUSY for EM models.
 Section~\ref{secondorder} is devoted to
 SSUSY. After getting compact expressions for
 intertwined Hamiltonians in terms of superpotentials and mass
 function, we obtain the solution of zero mode equation of
 supercharges. In Section~\ref{standarsusy} we study in detail the connection
 of the second order scheme with the standard one.
  Section~\ref{higherorder} contains the main result concerning SI for SSUSY scheme. A new relation between potential and
  mass function leading to SI model is derived. Two examples satisfying such relation are provided.
  Finally, Section~\ref{conclusion} contains the concluding remarks.

\section{First order SUSY and EM models}\label{firstorder}

The well-known superalgebra defined by the super Hamiltonian $H_s$
and the supercharges $Q, Q^{\dagger}$ is
\begin{equation}\label{supalg}
  H_s=\{Q,Q^{\dagger}\}, \  Q^2=Q^{\dagger^2}=
  [Q,H_s]=[Q^{\dagger},H_s]=0 ,
\end{equation}
where the symbol `$\dagger$' denotes the usual hermitian
conjugation.
 The supercharges are
represented in one dimensional quantum mechanics by the following
$2\times 2$ matrices
\begin{equation}\label{supch}
  Q=\left (
   \begin{array}{cc}
      0 & 0 \\
      \A & 0
   \end{array} \right ), \qquad
  Q^{\dagger}=\left (
   \begin{array}{cc}
      0 & \A^{\dagger} \\
      0 & 0
   \end{array} \right )\, .
\end{equation}
Consequently, the super Hamiltonian $H_s$ is diagonalized as
\begin{equation}\label{suph}
  H_s=\left (
    \begin{array}{cc}
      H_+ & 0 \\
      0 & H_-
    \end{array} \right ), \quad
  H_+=\A^{\dagger}\A, \quad H_-=\A\A^{\dagger}.
\end{equation}
In the standard SUSY approach to EM models the ladder operators
$\A,\A^{\dagger}$ are first order differential operators
\footnote{The representation of ladder operators is not unique for
EM models, for other variants see \cite{dut2,bag}.}
\begin{equation}\label{lad}
 \hspace{-2cm}\A =\frac{1}{\sqrt{m(x)}}\ \partial+\W(x)\, , \quad
 \A^{\dagger}=-\frac{1}{\sqrt{m(x)}} \ \partial+\W(x)-\left [ \frac{1}{\sqrt{m(x)}}\right ]'\, ,
   \qquad \left(\partial\equiv\frac{d}{dx}\right) \\
\end{equation}
where $\W(x)$ is the EM superpotential and the prime denotes
derivative with respect to $x$. Note that we have omitted the
factor $\hbar^2$ by defining the atomic units such that
$\hbar^2=2$. Denoting $H_{\pm}$ as
\begin{equation}\label{sch}
  H_{\pm}=-\partial \left( \frac{1}{m(x)}\ \partial\right)+V_{\pm}(x),
\end{equation}
the realization (\ref{lad}) straightforwardly expresses the
Schr\"odinger EM potentials in terms of mass and EM superpotential
\begin{equation}\label{scp}
\hspace{-2cm} V_+(x) =\W^2- \left [\frac{\W(x)}{\sqrt{m(x)}}\right
]'\, , \quad
    V_-(x)=V_+(x)+2\frac{\W'(x)}{\sqrt{m(x)}}-\frac{1}{\sqrt{m(x)}}
                     \left [\frac{1}{\sqrt{m(x)}}\right ]''.
\end{equation}

The main feature of above construction is the intertwining
relations between the bosonic and fermionic parts, which plays a
crucial role in SUSY theory
\begin{equation}\label{intw}
  \A H_+=H_-\A, \qquad H_+\A^{\dagger}=\A^{\dagger}H_- \, .
\end{equation}
The vacuum states may be characterized as
\begin{equation}\label{vac}
  \A \psi_0^+(x)=0, \qquad \mbox{or}\qquad \A^{\dagger}\psi_0^-(x)=0,
\end{equation}
where $\psi_n^{\pm}, n=0,1, \ldots$ denote the bound state wave
functions for the Schr\"odinger Hamiltonian $H_{\pm}$. Let us
point out that physically the mass function $m(x)$ is positive
definite and finite everywhere in the domain of definition of the
EM Schr\"odinger equation. Now, the existence of a zero mode will
depend on the asymptotic nature of the superpotential $\W(x)$ and
mass function $m(x)$. The ground state wave functions may be
computed from equation (\ref{vac})
\begin{equation}\label{gr}
 \hspace{-2cm}\psi_0^+(x) \propto \exp \left [-\int^x\sqrt{m(\tau)}\W(\tau)d\tau\right ] \, , \quad
       \psi_0^-(x) \propto \sqrt{m(x)}\exp \left
            [\int^x\sqrt{m(\tau)}\W(\tau)d\tau\right ] \, .
\end{equation}
The existence of the vacuum states is ensured if and only if the
integrals satisfy the following asymptotic condition
\begin{equation*}
  \int^x\sqrt{m(\tau)}\W(\tau)d\tau
        \begin{array}{c}
       \hspace{.1cm} \\[-1.2ex]
       \longrightarrow \\[-1.2ex]
       \scriptstyle x\rightarrow\pm \infty
       \end{array}
   +\infty\,(\mbox{ or}-\infty)\, .
\end{equation*}
Clearly both of the above conditions can not exist simultaneously
 and so at most one of the zero modes (\ref{gr}) will be
 annihilated in which case SUSY is unbroken (or exact). On the
 other hand, if neither of the zero modes exist then SUSY is called
 spontaneously broken. In this context it should be kept in mind that
 this argument fails for periodic Hamiltonians \cite{dun} because in that
 case one usually considers Bloch solutions and so the square-integrability
 criteria disappears [35--37]. In fact the question of existence of zero modes
 are related with the specified functional space on which a quantum system
 is to be considered \cite{tan3}. However in this treatise we will not consider
 periodic models. Thus for the systems defined on linear space it must be safely
 concluded that two zero modes can not exist simultaneously in the first order
 SUSY formalism. In the next section we will show that this
 is not the case for the SSUSY scheme.

For definiteness, let us suppose that the bosonic sector ($H_+$)
is fully known and possesses normalizable zero-energy state. Thus,
if either $\psi_0^+$ or $\W(x)$ is known, the other can be
obtained exactly by formulae (\ref{gr}). Now, one can extract full
knowledge about the fermionic sector ($H_-$) by the use of the
intertwining relations (\ref{intw}):
\begin{equation}\label{rw}
 \hspace{-2cm}\sqrt{E_{n+1}^+}\: \psi_n^-(x)=\A \psi_{n+1}^+(x)\, ,
 \sqrt{E_{n}^-}\:\psi_{n+1}^+(x)=\A^{\dagger} \psi_{n}^-(x)\, ;\quad
 E_n^-=E_{n+1}^+\, , E_0^+=0\, ,
\end{equation}
 for $n=0,1,2 \ldots$.

 Let us now consider the EM eigenvalue equation
\begin{equation}\label{eigen1}
 H_{\textrm{\tiny{EM}}}(x)\psi (x)\equiv [T_{\textrm{\tiny{EM}}}(x)+V_{\textrm{\tiny{EM}}}(x)]\psi (x)=\epsilon \psi (x)
\end{equation}
The first step to study this equation is certainly to choose a
suitable form of the hermitian kinetic energy operator
$T_{\textrm{\tiny{EM}}}(x)$. There is an intrinsic ambiguity in
selecting such form as this class of physical problems are
suffered from non-commutativity of momentum operator
$p=-i\sqrt{2}\partial$ and the effective mass operator $m(x)$.
Several forms had been proposed in the literature for
$T_{\textrm{\tiny{EM}}}(x)$, and considerable efforts were made to
remove the non-uniqueness of the kinetic energy operator [38--43].
But still the problem of ambiguity remains a open question in this
field. However one may rely on almost general representation
suggested in Ref.~\cite{von}
\begin{equation}\label{kin}
 T_{\textrm{\tiny{EM}}}(x)=\frac{1}{4} \left(m^{\textrm{a}}p\, m^{\textrm{b}}p\, m^{\textrm{c}}+m^{\textrm{c}}p\, m^{\textrm{b}}p\, m^{\textrm{a}}\right)\, ,
                      \quad   \textrm{a}+\textrm{b}+\textrm{c}=-1\, .
\end{equation}
The parameters $\textrm{a},\textrm{b},\textrm{c}$ in the above
equation are usually called `ordering parameters'. Most of the
present authors prefer to start from (\ref{kin}) as this
representation includes many special forms used in different
context. For instance, the authors in Ref.~\cite{plas} considered
following kinetic energy operator to apply first order SUSY
\begin{equation*}
 T_{\textrm{\tiny{EM}}}(x)=\frac{1}{2}\, p\left ( \frac{1}{m}\right )p\, ,
\end{equation*}
which was first proposed by BenDaniel and Duke \cite{ben}, and is
contained in the general representation (\ref{kin}) for the
special choice $\textrm{a}=\textrm{c}=0,\textrm{b}=-1$. In this
connection we would like to mention an interesting work
\cite{dut2} where the authors have proposed a new representation
of first order SUSY ladder operators including ambiguity
parameters of the kinetic energy operator (\ref{kin}) and have
obtained a substantial generalization over the result of
Ref.~\cite{plas}. It is now well-known that \cite{bag,ques} the
representation of ladder operators can be made free from the
ambiguity parameters [see equation (\ref{lad})] at the cost of
constraining the local potential $V_{\textrm{\tiny{EM}}}(x)$ in
equation (\ref{eigen1}) with a pseudo potential term $\rho(m)$
thereby considering the
 so-called effective potential energy $\widetilde{V}_{\textrm{\tiny{EM}}}(x)$.
For the two-parametric representation (\ref{kin}) of the kinetic
energy operator this pseudo potential term is given by
\begin{equation}\label{rho}
  \rho(m)=\frac{1+\textrm{b}}{2}\frac{m''}{m^2}-\eta\frac{m'^2}{m^3}\, , \quad
  \eta=1+\textrm{b}+\textrm{a}(\textrm{a}+\textrm{b}+1)\, ,
\end{equation}
where we have absorbed the parameter $\textrm{c }$ using the
constraint $\textrm{a}+\textrm{b}+\textrm{c}=-1$. In what follows
we will consider the following general EM Schr\"odinger equation
\begin{equation}\label{ems}
  \hspace{-2cm}H_{\textrm{\tiny{EM}}}(x)\psi(x)\equiv
  \left [-\partial(\frac{1}{m(x)}\partial)+\widetilde{V}_{\textrm{\tiny{EM}}}(x)\right ]\psi(x)
            =\epsilon\psi(x)\, ,\quad \widetilde{V}_{\textrm{\tiny{EM}}}(x)=V_{\textrm{\tiny{EM}}}(x)+\rho(m)\, .
\end{equation}
In above equation, $V_{\textrm{\tiny{EM}}}(x)$ represents the
local potential strength and the pseudo potential $\rho(m)$ is
given by (\ref{rho}), the latter depends on the ordering
parameters $\textrm{a},\textrm{b}$. We will assume that
$\textrm{a},\textrm{b}$ are real. The EM Hamiltonian (\ref{ems})
may then be identified with bosonic partner Hamiltonian $H_+$ in
(\ref{sch}) as
\begin{equation}\label{just}
  V_+(x)=\widetilde{V}_{\textrm{\tiny{EM}}}(x)-\epsilon\, , \quad \epsilon\leq  E_0^+=0\, ,
\end{equation}
by assigning $V_{\textrm{\tiny{EM}}}(x)=V_+(x)-\rho(m)+\epsilon$.
The quantity $\epsilon$ is often termed as factorization energy in
the SUSY procedure \cite{suk}.

\section{Second order SUSY and zero mode equation}\label{secondorder}

We will now  replace the intertwining operators $\A,\A^{\dagger}$
in equations (\ref{supalg})--(\ref{lad}) by the following second
order operators
\begin{equation}\label{hlad}
 \hspace{-2cm}\as=\frac{1}{m}\ \partial^2+\ws(x)\, \partial+c(x)\, ,
  \quad \as^{\dagger}=\frac{1}{m}\ \partial^2
   -\left [\ws(x)+ 2\frac{m'}{m^2}\right ] \partial+\left [ c(x)-\ws'_m(x)\right ] \, ,
\end{equation}
in which we have used the abbreviation
\begin{equation}\label{def2}
   \ws_m(x)=\ws(x)+\frac{m'(x)}{m^2(x)}\, .
\end{equation}
In the following we will use the terminology ``superpotential''
for the function $\ws_m(x)$. Clearly the super Hamiltonian $H_s$
given by (\ref{supalg})--(\ref{supch}), is now a fourth order
differential operator, and it will be physically meaningful if it
can be expressed as a quadratic polynomial in the usual EM
Schr\"odinger Hamiltonians. Thus we will introduce
\begin{equation}\label{emh}
 h_s=\left (
    \begin{array}{cc}
      h_+ & 0 \\
      0 & h_-
    \end{array} \right ), \quad
  h_{\pm}=-\partial\left(\frac{1}{m}\partial\right)+v_{\pm}(x).
\end{equation}

Our task is to determine the following matrix identity
\begin{equation}\label{id}
  H_s=h_s^2+l_1h_s+l_2 I_2\, ,
\end{equation}
where $l_1,l_2$ are arbitrary fixed real numbers. After some
involved but straightforward steps one may express $v_{\pm}(x)$ in
terms of $c(x)$, the superpotential $\ws_m(x)$ and the mass
function $m(x)$
\begin{equation}\label{res1}
v_{\pm}(x) =(\frac{1}{2}\mp
   1)\ws'_m\mp\frac{m'}{2m}\ws_m+\frac{m\ws_m^2}{2}-c(x)-\frac{l_1}{2}\, ,
\end{equation}
where the function $c(x)$ is given by
 \begin{equation}\label{c}
 \hspace{-2cm}c(x)=\frac{\ws_m'}{2}+\frac{m\ws_m^2}{4}-\frac{\ws''_m}{2m\ws_m}
 +\frac{1}{m}\left ( \frac{\ws'_m}{2\ws_m}\right )^2+\frac{3m'^2}{4m^3}
          -\frac{m''}{2m^2}-\frac{1}{m}\left (\frac{K_3}{2\ws_m}\right )^2.
 \end{equation}
 In above equations the quantity $K_3$ is taken, without loss of
generality as
 \begin{equation}\label{k3}
 K_3=+\sqrt{l_1^2-4l_2} .
 \end{equation}
 It should be emphasized that the quantity $K_3$ may be purely
 real or purely imaginary according as $l_1^2\ge 4l_2$ or $l_1^2<4l_2$. It
 is that constant which plays the role of determining reducible or
 irreducible SSUSY \cite{and3}. In the next section we will show
 that for hermitian quantum mechanics, SSUSY scheme could be
 reduced to first order SUSY for real $K_3$ only.

Note that the EM Hamiltonian $h_s$ always commutes with super
Hamiltonian $H_s$ due to the relation (\ref{id}), and hence both
have simultaneous eigenstates. Thus the intertwining relation
(\ref{intw}) implies that
 Schr\"odinger Hamiltonian (\ref{emh}) is doubly
 degenerated (up to zero modes of the supercharges), and wave functions are
 connected according to (\ref{rw}), where $E_n^{\pm}$ now denote eigenvalues of $h_{\pm}$.
 It should be mentioned that above expressions for $v_{\pm}$ and $c(x)$ satisfy
 additional intertwining relations between EM Hamiltonians $h_{\pm}$
\begin{equation}\label{int2}
  \as h_+=h_-\as\, , \qquad h_+\as^{\dagger}=\as^{\dagger}h_-\, .
\end{equation}
This relation is important not only because of its elegant
description, but also one can start from the requirement
(\ref{int2}), and may obtain expressions (\ref{res1}) and
(\ref{c}).

 We will now show that zero modes of both operators (\ref{hlad}) may exist simultaneously, in contrast to the standard first-order SUSY.
   To understand this let us write down the normalizable solutions
 of zero mode equations (\ref{vac}) for supercharges. Note that the  equation
 (\ref{vac}) becomes a second order differential equation with the replacement
 of $\A,\A^{\dagger}$ by $\as,\as^{\dagger}$ of (\ref{hlad}). These equations may be brought to
 the form similar to CM Schr\"odinger equation
\begin{equation}\label{sc}
 -\phi''(x)+\left [ \left (\frac{\ws_m'+K_3}{2\ws_m}\right )
 ^2+\left (\frac{\ws_m'+K_3}{2\ws_m}\right )'\right ]\phi(x)=0\, ,
\end{equation}
 by the transformations
 \begin{equation}\label{grh}
  \psi_0^{\pm}(x)= \sqrt{m(x)}\phi(x)
   \  e^{[\mp \frac12 \int^x m(\tau)\ws_m(\tau) d\tau ]}\, .
\end{equation}
The normalizability of $\psi_0^{\pm}$ clearly depends on
functional forms of both superpotential
  $\ws_m(x)$ and mass function $m(x)$. To illustrate it, consider that $\phi(x)$ in (\ref{sc})
  is identified with normalizable ground state wave function of a known solvable CM Hamiltonian
  by suitably choosing the superpotential $\ws_m(x)$. Then the prefactor $\phi(x)$ of $\psi_0^{\pm}$ in
  (\ref{grh}) is well-behaved, and hence both $\psi_0^{\pm}$ will be normalizable if $m(x)$ is also
  well-behaved and the integral $\int^xm\ws_m d\tau$ is finite. Thus in this instance
  both operators (\ref{hlad}) have normalizable zero modes, as was the
 situation for CM models. The explicit
 solutions for zero-mode states of both supercharges may be
 written as follows
\begin{equation}\label{gr222}
 \hspace{-2cm}\psi_{0,j}^{\pm}(x)\propto \sqrt{m\ws_m} \ e^{[\mp\int^x F_j(\tau)d\tau]}\, ,
 \: \: F_{j}(x)=\frac{m\ws_m^2+(-1)^j K_3}{2\ws_m}\, , \qquad j=1,2\, .
\end{equation}

Note that for $K_3=0$, $F_{1}(x)\equiv F_{2}(x)$ and so in that
case both operators may have at most one zero mode. It should be
mentioned that both of the zero modes are also formal eigenstates
of Schr\"odinger
 Hamiltonians $h_{\pm}$ for real $K_3$ only :
\begin{equation}\label{formal}
 h_{\pm}\psi_{0,j}^{\pm}=\eta_j\psi_{0,j}^{\pm}\, ,\qquad \eta_{j}=-\frac{l_1+(-1)^j K_3}{2}\, .
\end{equation}

Before concluding the section it may be pointed out that the
quantum systems built upon $\mathcal{N}$-th order representation
of ladder operators were categorized as type A $\mathcal{N}$-fold
SUSY (see for details Refs.\ \cite{tan3,tan1}). Hence it is not
difficult to show that the systems investigated in this article
are generically special cases of type A 2-fold SUSY. Furthermore
in Ref.~\cite{tan3} it was shown systematically that the zero
modes of one higher order supercharge can admit both physical and
non-physical states, the latter may be used as a good
transformation function to develop new solvable system \cite{as7}.

\section{Relation with Standard SUSY}\label{standarsusy}

In the previous section we have derived the expressions for
second-order SUSY partner potentials $v_{\pm}$. In this section we
want to investigate whether $v_{\pm}$ may be expressed through
standard SUSY formalism. That is, to say whether there exist
superpotentials $\W_1,\W_2$ in terms of which $v_{\pm}$ may be
brought to the first order form given by (\ref{scp}). To proceed
systematically, let us write $h_{\pm}$ as
\begin{equation}\label{pr} 
\hspace{-2cm} h_-\equiv h_-^{(1)}=\A_1\A_1^{\dagger}+K_1\, , \quad
 h_+\equiv h_+^{(2)}=\A_2^{\dagger}\A_2+K_2\, ,\quad A_j=\frac{1}{\sqrt{m}}\ \partial+\W_j\, ,
 \end{equation}
 where $\A_j$ are first order EM ladder operators and $K_1,K_2$ are suitable constants to be determined.
 In this section we will use the notations $h_-\equiv h_-^{(1)}$ and $h_+\equiv h_+^{(2)}$
 interchangeably for 2-SUSY partners $h_{\pm}$. Let us
 consider that $K_3$ is purely real. Assuming for definiteness $K_3>0$, two types
 of solutions for $\W_1,\W_2$ are possible.

\begin{itemize}
 \item[(a)]{\textbf{Type I:}}
 $\mathbf{K_1=-\frac{K_3+l_1}{2}\, , K_2=\frac{K_3-l_1}{2}}$

 The first order superpotentials are given by
 \begin{equation}\label{t1}
 \W_{1,2}=\left [ \frac{\ws_m\sqrt{m}}{2}+\frac{1}{2}\left
 (\frac{1}{\sqrt{m}}\right )'\right ]
 \pm \left
 [\frac{\ws_m'+K_3}{2\ws_m\sqrt{m}}-\frac{1}{2}\left (
 \frac{1}{\sqrt{m}}\right )'\right ] .
 \end{equation}
 The relation between second order operators $\as,\as^{\dagger}$
 and the first order operators $\A_1,\A_2$ in this case is
\begin{equation}\label{fs1}
 \as^{\dagger}\as=(\A^{\dagger}_2\A_2)(\A^{\dagger}_2\A_2+K_3) ,
 \quad \as\as^{\dagger}=(\A_1\A^{\dagger}_1)(\A_1\A^{\dagger}_1-K_3) .
\end{equation}

 \item[(b)]{\textbf{Type II:}}\
 $\mathbf{K_2=K_1=-\frac{K_3+l_1}{2}}$
 \begin{equation}\label{t2}
 \hspace{-2cm}\W_{1,2}=\left [ \frac{\ws_m\sqrt{m}}{2}+\frac{1}{2}\left
 (\frac{1}{\sqrt{m}}\right )'+\frac{K_3}{2\ws_m\sqrt{m}}\right ]
 \pm \left  [\frac{\ws_m'}{2\ws_m\sqrt{m}}-\frac{1}{2}\left (
 \frac{1}{\sqrt{m}}\right )'\right ] .
 \end{equation}
 The operator relation (\ref{fs1}) becomes
 \begin{equation*}
 \as^{\dagger}\as=(\A^{\dagger}_2\A_2)(\A^{\dagger}_2\A_2-K_3)\, ,
 \quad \as\as^{\dagger}=(\A_1\A^{\dagger}_1)(\A_1\A^{\dagger}_1-K_3) .
 \end{equation*}

\end{itemize}

 We see that both types of reductions are distinct for the superpotential $\W_2$ unless
 $K_3=0$. One may check readily that Type I reduction allows
 factorization of ladder operator $\as$ in terms of first order operators
 $\A_1,\A_2$ as
 \begin{equation*}
 \as=\A_1\A_2\qquad (\mbox{follows from the identity: }\A_1^{\dagger}\A_1=\A_2\A_2^{\dagger}+K_3\, ).
 \end{equation*}
 However this is not possible
 for Type II reduction. But in both cases $v_{\pm}$ are in 1-SUSY form,
 which is actually important. Their respective 1-SUSY partners may
 be written down using the formula (\ref{scp})
 \begin{equation}\label{part0}
 \hspace{-2.3cm}v_+^{(1)}=\W_1^2-\left (\frac{\W_1}{\sqrt{m}}\right )'+K_1\, ,
 \quad v_-^{(2)}=\W_2^2+\frac{\W_2'}{\sqrt{m}}-\W_2\left
 (\frac{1}{\sqrt{m}}\right )'-\frac{1}{\sqrt{m}}\left (\frac{1}{\sqrt{m}}\right )''+K_2\, .
 \end{equation}
 According to standard SUSY, $v_{\pm}^{(1)}$ and $v_{\pm}^{(2)}$
 are isospectral where $v_-^{(1)}\equiv v_-$ and $v_+^{(2)}\equiv
 v_+$ are given by (\ref{res1}). Hence we see that second order
 SUSY formalism may give us opportunity of studying two standard SUSY
 pairs $(h_+^{(j)},h_-^{(j)})$ simultaneously for $j=1,2$. The ground states of both pairs of Hamiltonian are zero modes of first
order operators
 $\A_j,\A_j^{\dagger}$ given by (\ref{pr}). These  can be computed by substituting for $\W_1,\W_2$ from (\ref{t1}) and (\ref{t2})
 (corresponding to Type I and Type II reduction) into the formulae (\ref{gr}) for $\W$. For instance the
zero modes for
 the opertors $\A_j,\A_j^{\dagger}, j=1,2,$
 corresponding to Type II reduction are
\begin{equation}\label{grrn02}
 \hspace{-2cm}\phi_{0}^{+(1)}\propto \frac{e^{-\int^x F_2(\tau)d\tau}}{\sqrt{\ws_m}} \, ,  \phi_{0}^{-(1)}=\psi_{0,2}^- \, ;\qquad
 \phi_{0}^{-(2)}\propto\frac{e^{+\int^x F_2(\tau)d\tau}}{\sqrt{\ws_m}}\, , \phi_{0}^{+(2)}=\psi_{0,2}^+ \, ,
\end{equation}
where $\psi_{0,2}^{\pm}$ are given by (\ref{gr222}) and
$\phi_0^{+(j)},\phi_0^{-(j)}$ denote ground state wave functions
for the Hamiltonians $h_+^{(j)},h_-^{(j)}$ for $j=1,2$
respectively.

Let us point out that for imaginary $K_3$, given by (\ref{k3}),
the above reduction is not possible in hermitian quantum mechanics
\cite{sam2}, because both first order superpotentials $\W_1,\W_2$
will be complex [see equations (\ref{t1}) and (\ref{t2})]. In fact
this will lead us to an irreducible transformation between real
and complex potentials for EM Hamltonians \cite{as11}.

The discussion in this section clearly shows that the
factorization of a higher order linear differential operator in
terms of lower order operators is in general non-unique. That is
to say, one operator can admit both reducible and irreducible
representations. Hence from strictly mathematical sense the
concept of reducibility of HSUSY must be defined on the basis of
an additional restriction on Hamiltonians that they are
factorizable according to (\ref{t1}) or (\ref{t2}). In the next
section we will propose a higher order SI criteria for EM
Hamiltonians. Our purpose of introducing this section is to
compare the result of SI obtained through HSUSY with that obtained
via first order SUSY. This will give us a better insight about why
and how SI scheme proposed in this article is an important
generalization over the SI formalism in the standard approach.

\section{Higher order SI criteria for EM Hamiltonian}\label{higherorder}

In Ref.~\cite{plas} two types of SI criteria were discussed for EM
Hamiltonian generated through first order SUSY let apart the
generalized treatment proposed in Ref.~\cite{dut2}. Recently a
kind of deformed SI criteria has been introduced \cite{bag}, which
is also in the context of standard SUSY. Searching for SI
Hamiltonian is useful, because this integrability condition leads
to certain relation between potential and mass functions producing
an ES model. For instance, claiming first order SUSY partners
$V_-(x;\lambda)-V_+(x;\lambda)=2\lambda$, from equation
(\ref{scp}), one obtains following relation between first order
superpotential $\W (x)$ and mass $m(x)$ :
\begin{equation*}
  \W(x)=\frac{1}{2}\left (\frac{1}{\sqrt{m}}\right )'+\lambda
  \int^x \sqrt{m}d\tau \, .
\end{equation*}
Just this relation was discovered in Ref~\cite{plas}. Here we wish
to enquire this simple SI condition for SSUSY partners
$(v_+,v_-)$, given by (\ref{res1}).

\subsection{Theoretical construction}\label{theory}
Let us consider
\begin{equation}\label{si}
  h_-(x;\lambda)=h_+(x;\lambda)+2\lambda\, ,  \qquad \lambda>0\, .
\end{equation}
It is not very difficult to see that the above requirement
expresses second order superpotential in the form
\begin{equation}\label{sup}
  \ws_m(x)=\frac{g(x)}{\sqrt{m(x)}}\, ,\quad
  g(x)=\gamma+\lambda\int^x\sqrt{m}d\tau\, ,
\end{equation}
where $\gamma$ is an integration constant. Substituting this
expression for $\ws_m(x)$ into (\ref{res1}) for $v_+(x)$, we
obtain following relation between potential and mass
\begin{equation}\label{ssi}
  v_+(x;\lambda)=\frac{g^2}{4}-\frac{\lambda^2-K_3^2}{4g^2}+\frac{m''}{4m^2}-\frac{7}{16}\frac{m'^2}{m^3}
    -\left (\lambda+\frac{l_1}{2}\right )\, ,
\end{equation}
where we are considering $K_3>0$.

We stress that this is a new relation between potential and mass
which gives us a class of SI Hamiltonian $h_+$. The whole set
$\{\psi_n^+(x;\lambda),E_n^+\}$ of eigenstates and spectra for
$h_+$ can be constructed by exploiting  the intertwining relation
(\ref{int2}). The procedure is, as in the case for harmonic
oscillator, to apply successively the raising operator
$\as^{\dagger}$ upon zero mode of the lowering operator $\as$.
Note that the second order operator $\as$ has two zero modes
$\psi_{0,j}^+\, ,j=1,2$, given by (\ref{gr222}). Both these zero
modes are formal solutions of Schr\"odinger equation for $h_+$
[see equation (\ref{formal})] with the eigenvalues
$\eta_1=(-l_1+K_3)/2$ and $\eta_2=-(l_1+K_3)/2$ , where
$\eta_1>\eta_2$. Hence we obtain double sequences of eigenstates
based on both zero modes. The labels of eigenstates will depend on
the values of the SI parameter $\lambda$ in equation (\ref{si}).
For simplicity let us first consider the case
\begin{equation}\label{range1}
\hspace{2cm}\lambda> K_3/2\, , \qquad K_3>0
\end{equation}
 In this case the wave functions and energy eigenvalues of $h_+(x;\lambda)$ may
 be expressed as follows
\begin{equation}\label{finwave}
 \hspace{-.5cm} \left . \begin{array}{llll}
     \psi_{2n}^+(x;\lambda) & =\left (\as^{\dagger}\right )^n \psi_{0,2}^+(x)\, , & E_{2n}^+ & =\eta_2+2n\lambda\, , \\
     \psi_{2n+1}^+(x;\lambda)& =\left (\as^{\dagger}\right )^n \psi_{0,1}^+(x)\, , & E_{2n+1}^+ & =\eta_1+2n\lambda\, ,
           \end{array}\right \}\, ,\: n=0,1,\ldots
\end{equation}

We will now turn to the case for $\lambda\leq K_3/2$. Note that
for all values of $\lambda (>0)$, the ground state will be given
by the zero mode $\psi_{0,2}^+$ with the eigenvalue $\eta_2$.
Higher excited states will be obtained in a similar procedure as
described above, but one has to relabel the states according to
the range of values of $\lambda$. To illustrate let us take the
most general situation where
\begin{equation}\label{range2}
 K_3/2(\kappa+1)<\lambda<K_3/2\kappa\, , \quad \kappa=1,2,\ldots
\end{equation}
 Here first $\kappa$ members will be given by the sequence based on
the zero mode $\psi_{0,2}^+$. Thus the wave functions  are
\begin{equation}\label{finwave2a}
 \psi_n^+(x;\lambda)=\left \{  \begin{array}{ll}
    \left (\as^{\dagger}\right )^n \psi_{0,2}^+(x)\, ,& n=0,1,\ldots \kappa \\
 \left (\as^{\dagger}\right )^{\frac{n+\kappa}{2}} \psi_{0,2}^+(x)\, ,& n=\kappa+2,\kappa+4,\ldots \\
 \left (\as^{\dagger}\right )^{\frac{n-\kappa-1}{2}} \psi_{0,1}^+(x)\, ,& n=\kappa+1,\kappa+3,\ldots
                               \end{array} \right .
\end{equation}
with the spectra
\begin{equation}\label{finwave2b}
 E_n^+=\left \{  \begin{array}{ll}
    \eta_2+2n\lambda\, ,& n=0,1,\ldots \kappa \\
  \eta_2+(n+\kappa)\lambda\, ,& n=\kappa+2,\kappa+4,\ldots \\
 \eta_1+(n-\kappa-1)\lambda\, ,& n=\kappa+1,\kappa+3,\ldots
                               \end{array} \right .
\end{equation}
It is a trivial exercise to verify that for $\lambda=K_3/2\kappa$,
the sequence built on $\psi_{0,1}^+$ coincides with that built on
$\psi_{0,2}^+$ except for the first $\kappa$ members, which are
the lowest states and only singular solutions. The wave functions
and spectra will be provided by single sequence
\begin{equation}\label{finwave3}
 \hspace{-1cm} \lambda=\frac{K_3}{2\kappa}:\qquad \psi_n^+ (x;\lambda)=\left (\as^{\dagger}\right )^n \psi_{0,2}^+(x)\, ,
  \quad E_n^+=\eta_2+n\frac{K_3}{\kappa}\, ,\: n=0,1,\ldots .
\end{equation}

It should also be kept in mind that for SSUSY scheme, zero modes
of both operators $\as$ and $\as^{\dagger}$ may simultaneously
exist. This implies that the above sequences may terminate at $n=N
(\geq 1)$ if $(\as^{\dagger})^N\psi^+_{0,j}\propto \psi_{0,j'}^-\,
, j,j'=1,2,$ $\psi_{0,j'}^-$ being the possible zero modes of
$\as^{\dagger}$, given by (\ref{gr222}). Note that for
$\lambda=K_3$, we have from (\ref{finwave})
harmonic-oscillator-like spectra
\begin{equation*}
  E_n=-\left ( \lambda+\frac{l_1}{2}\right )+\left (n+\frac{1}{2}\right )\lambda\, , \quad n=0,1,\ldots\, ,
\end{equation*}
while for $\lambda\neq K_3$, the spectra given by (\ref{finwave}),
(\ref{finwave2b}) or (\ref{finwave3}) separately coincide with
those of singular oscillator.

Under the SI condition (\ref{si}), all levels can be obtained in a
closed analytic form. For instance, first three members of the
sequence $\{\left (\as^{\dagger}\right )^n \psi_{0,2}^+(x)\}$ are
\begin{equation}\label{finwave1}
  \hspace{-2.5cm} \left . \begin{array}{l}
  \psi_{0,2}^+(x)\propto m^{1/4}(x)\left [g(x)\right ]^{\frac{\lambda- K_3}{2\lambda}}\textrm{e}^{-\frac{g^2(x)}{4\lambda}}\, ,
 \quad \as^{\dagger}\psi_{0,2}^+(x)\propto \left [g^2(x)+K_3-2\lambda\right ]\psi_{0,2}^+(x)\, , \\ [1ex]

 \left (\as^{\dagger}\right )^2 \psi_{0,2}^+(x)\propto \left [\left (g^2(x)+K_3-4\lambda\right )^2
                              +2\lambda\left (K_3-4\lambda\right )\right ]\psi_{0,2}^+(x)\, .
                   \end{array}\right \}
\end{equation}
Higher members can be constructed in a similar fashion. The
members of the other sequence $\{\left (\as^{\dagger}\right )^n
\psi_{0,1}^+(x)\}$ can be obtained by simply changing the sign of
$K_3$ in the corresponding members of the sequence $\{\left
(\as^{\dagger}\right )^n \psi_{0,2}^+(x)\}$. The nature of the
function $g(x)$ given by (\ref{sup}) is crucial for the
normalizability of the wave functions. In the first place, $m(x)$
must be so chosen that it verifies $g^2(x)\rightarrow +\infty$ as
$x\rightarrow \pm \infty$ and remains finite otherwise. Secondly,
suppose that $g(x)$ is nodeless in the whole domain. Then the
potential $v_+(x;\lambda)$ is non-singular and both sequences
provide regular (non-singular) solutions. But if $g(x)$ has a node
at $x=x_0$ then the potential is singular at that point for
$\lambda\neq K_3$. Let us note that the node of $g(x)$ is of first
order, because $g'(x)=\lambda\sqrt{m(x)}$ is always nodeless for
chosen mass function. Hence near $x\thicksim x_0$, $v_+(x)$ will
behave like $C(x-x_0)^{-2}$, where the constant $C$ characterizes
the strength of the singularity. It is well-known that
self-adjoint extension of such Hamiltonian can be determined on
the whole domain for the range $-1/4<C<3/4$. The singularity will
be attractive or
 repulsive according as $\lambda>K_3>0$ or $0<\lambda<K_3$. One
 may readily check from (\ref{finwave1}) that
$\psi_{0,2}^+(x)$ is singular for $\lambda<K_3$, and so in this
case depending on the values of $\lambda$ and other parameters
some or all of the members of the sequence $\{\left
(\as^{\dagger}\right )^n \psi_{0,2}^+(x)\}$ have to be deleted
from the set of regular solutions of $h_+(x;\lambda)$.

Hence we have proved that the Hamiltonian $h_+(x;\lambda)$ with
the potential $v_+(x;\lambda)$ in (\ref{ssi}) is a new ES model
possessing second order SI condition in the EM framework. Two
remarks are in order. As one lowers the value of $\lambda$ by
increasing the value of $\kappa$ in (\ref{range2}), the levels
become closer and closer and in the limit $\lambda\rightarrow 0
(\kappa\rightarrow\infty)$ only two levels will be left namely
$\psi_{0,2}^+(x)$ and $\psi_{0,1}^+(x)$ with the eigenvalues
$\eta_2,\eta_1$ respectively. Clearly this gives a quasi-exactly
solvable system with two known levels. Since in this article we
are only interested about ES models, $\kappa$ will be a finite
quantity or equivalently the SI parameter $\lambda$ is a strictly
non-zero finite positive number. Secondly we have already
mentioned that if the function $g(x)$ in (\ref{sup}) has a node at
$x=x_0$, the potential $v_+(x;\lambda)$ given by (\ref{ssi}) will
be singular at that point. Hence extreme care should be taken to
decrease (or increase) the value of $\lambda$ in order to keep the
strength $C$ of the singularity controlled ($-1/4<C<3/4$), as for
a given mass function $m(x)$ and the constant $K_3$ the strength
$C$ is inversely proportional with $\lambda$. This will be clear
if one expands the function $g(x)$ about its node $x_0$ giving
\begin{equation}\label{C}
  \hspace{2cm} C=\frac{\left (K_3/\lambda\right )^2-1}{4m(x_0)}\, .
\end{equation}
Next we are going to construct two classes of examples based on
the theoretical model proposed in this subsection, where this last
comment will be crucial to get the physically acceptable
Hamiltonians.

\subsection{Hyperbolic mass and potential}

Let us choose the mass function as
\begin{equation}\label{ex1}
 m(x)=(\alpha+\beta\tanh x)^2\, , \quad \beta\neq 0, |\alpha|> |\beta|\, .
\end{equation}
This mass depicts a smooth step function. It increases (or
decreases) from the value $(\alpha-\beta)^2$ for $x=-\infty$ to
the value $(\alpha+\beta)^2$ for $x=+\infty$ according as
$\alpha\beta>0$ (or $\alpha\beta<0$). From (\ref{sup}) the
function $g(x)$ may be computed
\begin{equation}\label{g1}
 g(x;\lambda)=\gamma+\lambda \left [\alpha x+\beta\ln \cosh x \right ]\, .
\end{equation}
The potential turns out to be
\begin{equation}\label{si2}
\hspace{-2cm}v_+(x;\lambda) =
\frac{g^2}{4}-\frac{\lambda^2-K_3^2}{4g^2}
  -\left (\lambda+\frac{l_1}{2}\right )
 +  \frac{\beta\textrm{sech}^2 x[\beta\tanh^2 x-4\alpha\tanh
x-5\beta]}{4(\alpha+\beta\tanh x)^4}\, .
\end{equation}
\begin{figure}[ht]
\centering \epsfig{file=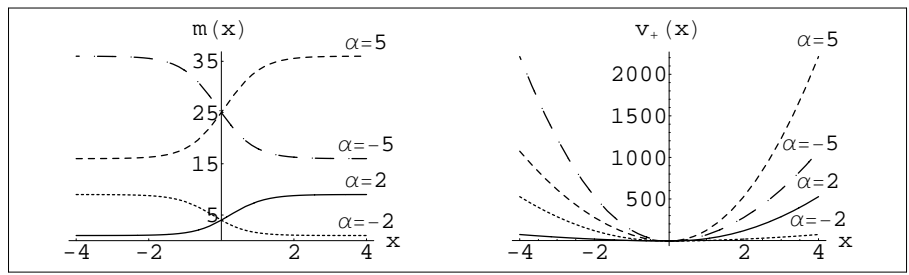, width=12cm}
 \caption{\small The effective mass (\ref{ex1}) and
potential function (\ref{si2}) with
$l_1=5,\beta=\gamma=1,\lambda=K_3=4$ that depicts harmonic
oscillator-like barrier. The graphs are drawn in the atomic units
(a.u.) defined by $\hbar^2=2$ for four different values of the
parameter $\alpha$.} \label{fig1}
\end{figure}
The nonsingular SI potential (\ref{si2}) and effective mass
(\ref{ex1}) are shown in Fig.~\ref{fig1} corresponding to (i)
$\beta=\gamma=1,l_1=5,\lambda=K_3=4$ and (ii) four different
values of the parameter $\alpha$. For all values of $\alpha$ (with
the restriction $|\alpha|>|\beta|,\beta\neq 0$, which is necessary
to have nodeless nonconstant mass) we have a harmonic
oscillator-like well. As the magnitude of $\alpha$ decreases the
well becomes flatter. The constant mass system will be recovered
with $\beta=0$, and consequently the potential will reduce to
harmonic oscillator for $\lambda=K_3$. For very small values of
$\beta$ the wave functions tend to those associated with harmonic
oscillator. On the contrary, for large values of $\beta$ the wave
functions tend to spread over the region whose size grows as
$\beta$ increases due to the influence of step mass (\ref{ex1}) in
the potential profile. This behaviour is demonstrated in
Fig.~\ref{fig2}, which contains the potential and wave functions
associated with ground and first two excited states corresponding
to (i) $\alpha=\gamma=1,\lambda=K_3=4$ and (ii) $\beta=0.9,0.001$.
\begin{figure}[ht]
\centering \epsfig{file=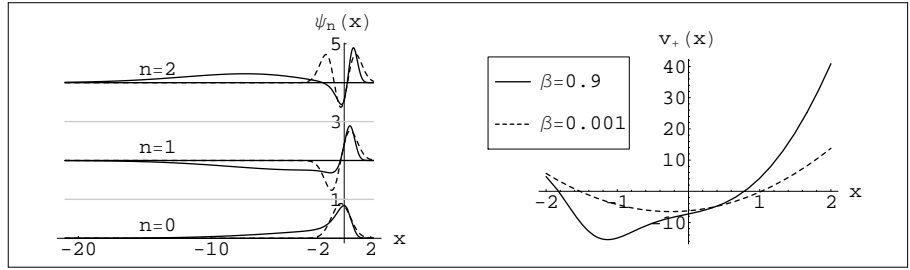, width=12cm}
 \caption{\small Wave functions associated with ground and first two excited states
and the SI potential (\ref{si2}) corresponding to $\beta=0.9$ and
$\beta=0.001$ while other parameters are
$\alpha=\gamma=1,\lambda=K_3=4$ (a.u. are used).} \label{fig2}
\end{figure}

\begin{figure}[ht]
\centering \epsfig{file=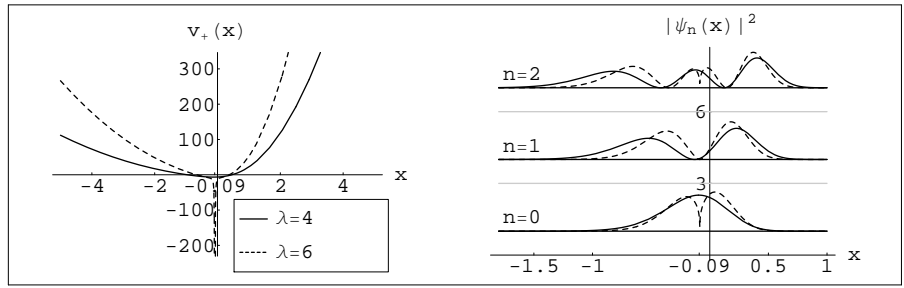, width=13.5truecm}
\caption{Non-singular and singular (attractive) potential
(\ref{si2}) along with probability-density functions (in a.u.)
associated with ground and first two excited states corresponding
to $\lambda=4$ and $\lambda=6$ for
$l_1=5,K_3=4,\alpha=2,\beta=\gamma=1$.} \label{fig3}
\end{figure}

The nature of the potential is dramatically changed for
$\lambda\neq K_3$. As can be seen from (\ref{g1}) that near
$x\thicksim x_0$, the potential behaves like
\begin{equation*}
  v_+(x;\lambda)\thicksim \frac{C}{(x-x_0)^2}\, ,
\end{equation*}
where $x=x_0$ is node of $g(x)$ and the strength $C$ of the
singularity is to be computed from the formula (\ref{C}). For the
weakly attractive singularity ($-1/4<C<0$) the physical wave
functions tend to vanish at $x\rightarrow x_0$ according to the
boundary conditions. This behaviour is clearly appreciated from
Fig.~\ref{fig3}, where we have plotted singular and non-singular
potentials with their probability density functions
($|\psi_n(x)|^2$) associated with first three lowest levels
corresponding to (i)$\lambda=4$ and (ii) $\lambda=6$. The other
constants are taken as $l_1=5,K_3=4,\beta=\gamma=1,\alpha=2$,
which gives $C=-0.04$ and $x_0=-0.09$.

\subsection{Algebraic mass and potential}

As our second example, let us take the following mass function
studied in Ref. \cite{plas}
\begin{equation}\label{ex2}
 m(x)=\left (\frac{\alpha+x^2}{1+x^2}\right )^2\, , \quad
 \alpha>0\, , \  \neq 1 .
\end{equation}
This mass function remains strictly positive definite everywhere
and approach a constant value 1 at both infinity. The SI potential
is
\begin{equation} \label{si1}
\hspace{-2cm}
 v_+(x;\lambda)=\frac{g^2}{4}-\frac{\lambda^2-K_3^2}{4g^2}-\left(\lambda+\frac{l_1}{2}\right )
 +(\alpha-1)\frac{3x^2-1}{(\alpha+x^2)^3}-5(\alpha-1)^2\frac{x^2}{(\alpha+x^2)^4}\, ,
\end{equation}
where
\begin{equation}\label{g2}
  g(x;\lambda)=\gamma+\lambda [x+(\alpha-1)\textrm{tan}^{-1}x]\, .
\end{equation}%

\begin{figure}[ht]
\centering \epsfig{file=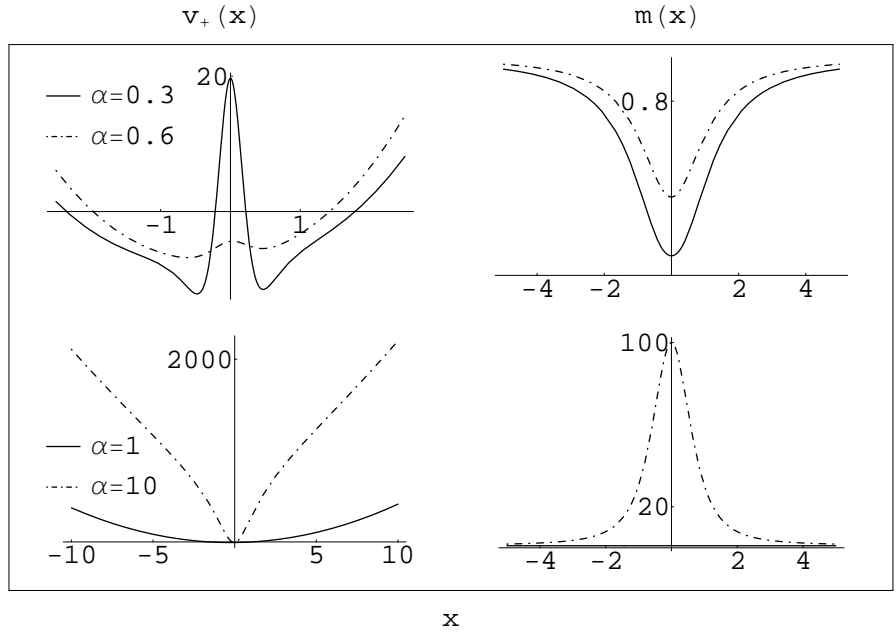, width=13.8cm}
 \caption{\small Non-singular potential (\ref{si1}) and mass function (\ref{ex2}),
depicted in a.u., for $l_1=5,\gamma=1,\lambda=K_3=4$ corresponding
to (i) $\alpha=0.3,0.6$ (first row) and (ii) $\alpha=1,10$ (second
row).} \label{fig4}
\end{figure}%

 The non-singular potential is given by
(\ref{si1}) for $\lambda=K_3$. For $0<\alpha<1$, it is a bistable
potential and for $\alpha>1$ it is a single potential well, while
$\alpha=1$ recovers the constant-mass system giving harmonic
oscillator well. Fig.~\ref{fig4} describes non-singular potential
(\ref{si1}) and mass function (\ref{ex2}) corresponding to (i)
$\alpha=0.3,0.6$ (first row) and (ii) $\alpha=1,10$ (second row).
From this Fig.\ it is clear that for small values of $\alpha$, the
height of separator between two wells grows producing a thin
barrier. On the other hand for very large values of $\alpha$, the
single well becomes sharper compare to that of standard harmonic
oscillator well.

\begin{figure}[ht]
\centering \epsfig{file=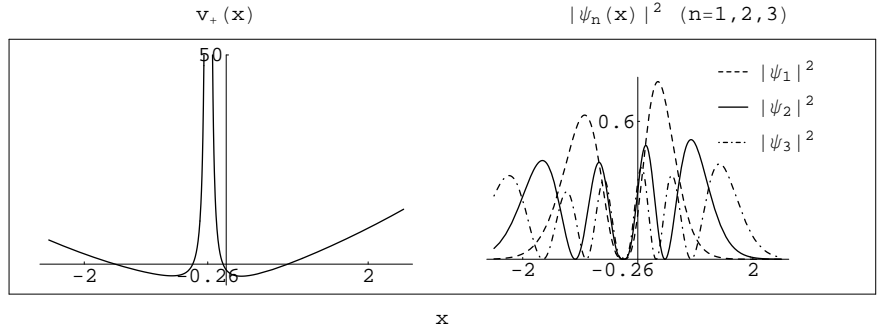, width=15cm}
 \caption{\small Singular (repulsive) potential (\ref{si1}) and the probability-density
functions (in a.u.) corresponding to non-singular wave functions
associated with first three excited levels for
$l_1=5,\alpha=2,\gamma=1,\lambda=2, K_3=4$.} \label{fig5}
\end{figure}

The potential acquires an inverse-square singularity for
$\lambda\neq K_3$, which is repulsive (attractive) for
$0<\lambda<K_3$ ($\lambda>K_3$). For weakly repulsive singularity
($0<C<3/4$) the regular (non-singular) wave functions are given by
the sequence $\{\left (\as^{\dagger}\right )^n \psi_{0,1}^+(x)\}$.
 According to the boundary conditions required to have self-adjoint
extensions of the Hamiltonian, these wave functions tend to vanish
at $x\rightarrow x_0$, $x_0$ being node of the function $g(x)$
given by (\ref{g2}). This behaviour is depicted in
Fig.~\ref{fig5}, wherein we have plotted singular potential along
with probability density functions associated with first three
excited levels of the sequence $\{\psi_{n}(x;\lambda)\}$ given by
(\ref{finwave3}) for
 $\lambda=2$, other parameters being taken as
$l_1=5,K_3=4,\alpha=2,\gamma=1$. These choice of parameters yields
the strength $C$
 and center $x_0$ of the singularity as $C=0.25,x_0=-0.26$.

Physically the well confining the particle is divided into two
zones by a high thin barrier at $x_0=-0.26$. But note that for
weakly repulsive case this barrier is penetrable by a
quantum-mechanical particle.

\subsection{SI condition in reduced first order SUSY}

 In subsection~\ref{theory} we have proposed a second order SI condition (\ref{si})
 and consequently have obtained a new system of ES Hamiltonians $h_+(x;\lambda)$
  where the potential $v_+(x;\lambda)$ and the mass $m(x)$ are related according to (\ref{ssi}).
  Here we wish to show that the same system (\ref{ssi})
 (up to a constant shift) could be studied through first order
 SI formalism provided the Hamiltonians are factorizable according to (\ref{t1})
 or (\ref{t2}). To see this, consider the 1-SUSY pair $(v_+\equiv
v_+^{(2)},v_-^{(2)})$ given by (\ref{res1}) and (\ref{part0})
corresponding to Type II reduction (\ref{t2}). Here the SI
parameter will be $K_3$ instead of $\lambda$, while the other
parameters have to be considered fixed. For convenience, in the
following we will use the symbol $a_0$ for $K_3$. The relation
between  first order superpotential $\W_2(x;a_0)$ and the mass
function $m(x)$ reads from (\ref{t2}) and (\ref{sup})
\begin{equation}\label{rsi}
 \W_2(x;a_0)=\frac{g(x)}{2}-\frac{m'(x)}{4m^{3/2}(x)}+\frac{a_0-\lambda}{2g(x)}\, .
\end{equation}
To have zero energy ground state, consider the Hamiltonian
$\tilde{h}_+^{(2)}(x;a_0)=h_+^{(2)}(x;a_0)-K_2(a_0)=\A_2^{\dagger}(x;a_0)\A_2(x;a_0)$,
where the operator $\A_2$ is given by (\ref{pr}) and
$K_2(a_0)=-(a_0+l_1)/2$ for Type II reduction. Let us use the
following abbreviations for the shifted potentials
\begin{equation*}
 \hspace{-2cm}\tilde{v}_+^{(2)}(x;a_0)=v_+^{(2)}(x;a_0)-K_2(a_0)\, ,
\qquad \tilde{v}_-^{(2)}(x;a_0)=v_-^{(2)}(x;a_0)-K_2(a_0)\, ,
\end{equation*}
where $\phi_n^{+(2)}(x;a_0),\epsilon_n^{+(2)}$ correspond wave
functions and eigenvalues. Note that the wave functions and energy
eigenvalues of $h_+(x;\lambda)$, given by (\ref{emh}) and
(\ref{ssi}) are related with
$\{\phi_n^{+(2)}(x;a_0),\epsilon_n^{+(2)}\}$ by
\begin{equation*}
  \psi_n^+(x;\lambda)\propto \phi_n^{+(2)}(x;a_0)\, ,
 \qquad  E_n^+=\epsilon_n^{+(2)}+K_2\, , \quad n=0,1,\ldots
\end{equation*}
where $a_0\equiv K_3$. One then find from (\ref{rsi}) following SI
condition in the reduced first order SUSY
\begin{equation}\label{rsis}
 \hspace{-2cm}\tilde{v}_-^{(2)}(x;a_0)=\tilde{v}_+^{(2)}(x;a_1)+R(a_1)\, ,
 \qquad a_1=f(a_0)=2\lambda-a_0\, , \: R(a_1)=2\lambda-a_1\, .
\end{equation}
Thus full spectra of $\tilde{h}_+^{(2)}(x;a_0)$ can be recovered
according to standard prescription
\begin{equation*}
\hspace{-.8cm} \left . \begin{array}{l}
 \epsilon_0^{+(2)}=0 , \epsilon_n^{+(2)}=\sum_{k=1}^{n}R(a_k)\, ,\quad a_k=f(a_{k-1})
                              \\ [1ex]
 \phi_n^{+(2)}(x;a_0)\propto \A_2^{\dagger}(x;a_0)\A_2^{\dagger}(x;a_1)\cdots
 \A_2^{\dagger}(x;a_{n-1})\phi_0^{+(2)}(x;a_n)
 \end{array}\right \}\, , \quad n=1,2\ldots
\end{equation*}
where $\phi_0^{+(2)}(x;a_0)$ is zero mode of the operator
$\A_2(x;a_0)$, given by (\ref{grrn02}). Two points are to be
noted. Firstly imposing the restriction of factorizability on the
Hamiltonians one would come out with the SI condition (\ref{rsis})
and the relation (\ref{rsi}), which have not been reported so far
in the literature for EM Hamiltonian. Secondly the proposed SSUSY
scheme definitely leads to a generalized SI criteria (\ref{si}),
because in this case no such restrictions need to be imposed on
the Hamiltonians.

Our final remark is that in the CM limit $m\rightarrow 1$
\footnote{$m\rightarrow 1$ is taken to tally with the conventional
choice of atomic units $\hbar^2=2m=1$ for constant mass $m$.},
both potentials $v_+(x;\lambda)$, given by (\ref{si2}) and
(\ref{si1})
 reduce to known SI potential $v_+^{\textrm{\tiny{CM}}}(x;\lambda)$ :
\begin{equation}\label{cmsi}
 v_+^{\textrm{\tiny{CM}}}(x;\lambda)=\frac{(\gamma+\lambda x)^2}{4}-\frac{\lambda^2-K_3^2}{4(\gamma+\lambda x)^2}
 -\left (\lambda+\frac{l_1}{2}\right )\, ,
\end{equation}
and the wave functions and energy eigenvalues exactly coincide
with those \cite{lev,as6} for (\ref{cmsi}).

\section{Conclusion}\label{conclusion}
 In this article we have used SSUSY scheme for describing dynamics
of a  quantum particle with a position-dependent mass. We have
derived a compact expression of 2-SUSY pairs $(v_+,v_-)$ in terms
of second order superpotential $\ws_m(x)$ and the mass function
$m(x)$. A detailed analysis has been given about zero mode
equations of second order supercharges and possible reduction of
SSUSY scheme to first order SUSY. Based on the existence of
intertwining relation between 2-SUSY partner Hamiltonians
$h_{\pm}$, we have obtained a new relation between potential
$v_+(x)$ and mass $m(x)$ leading to a simple SI condition. As a
result full spectra is achieved by successive application of
second order raising operator $\as^{\dagger}$ upon the zero modes
of the lowering operator $\as$. It is shown that in the reduced
first order SUSY approach, one may obtain a new relation between
first order superpotential $\W_2(x)$ and mass $m(x)$ and the same
system (up to a constant shift) could possess a first order SI
property provided the Hamiltonians are factorizable. The advantage
of using SSUSY scheme is that it deals with a generalized but
simpler SI requirement, namely partner Hamiltonians differ by a
constant $\lambda$.

We have constructed explicit examples for two types of
position-dependence of mass function, one is hyperbolic and the
other is algebraic. The corresponding potentials have very
different aspects based on the values of the SI parameter
$\lambda$. Both are non-singular for $\lambda=K_3$, where the
constant $K_3$ appears in the process of relating Schr\"odinger
Hamiltonian with super Hamiltonian. The parameters $\beta$ and
$\alpha$ characterizes position-dependence of mass (\ref{ex1}) and
(\ref{ex2}) respectively. The non-singular algebraic potential
(\ref{si1}) shares same qualitative features as those discovered
in Ref. \cite{plas}. For $\lambda\neq K_3$, the potentials acquire
an inverse square singularity, which is attractive (repulsive) for
$\lambda>K_3$ ($\lambda<K_3$). The week strength of singularity
($-1/4<C<3/4$) is particularly of physical interest. Because for
weakly attractive singularity ($-1/4<C<0$) the ground state energy
remains finite. On the other hand, for weakly repulsive
singularity ($0<C<3/4$) the barrier is penetrable by
quantum-mechanical particle. In both instances, regular
(non-singular) wave functions vanish at the center of the
singularity.

It is important to clarify that the novelty of our work lies in
the fact that we have generalized for the first time the concept
of SI to the second order SUSY approach for EM Hamiltonians. In
this context one should look into the explicit classification done
in Ref.~\cite{tan1} via $\mathcal{N}$-fold SUSY approach to EM
quantum systems. The systems investigated in this article belong
to a special subclass of the systems studied there, which possess
second order shape-invariance. Hence one has a significant
advantage of obtaining wave functions and energy eigenvalues for
such Hamiltonians over the existing general method for arbitrary
potential and mass functions.

We would like to mention that in many practical applications,
where continuous spectra is of interest \cite{basbk}, SSUSY scheme
used here may be utilized to relate transmission and reflection
amplitudes of partner Hamiltonians. The idea of SI condition for
SSUSY scheme can be straightforwardly extended to general
$\mathcal{N}$-th order representation ($\mathcal{N}>2$) of ladder
operators. However obtaining a physical model will be much more
difficult, because one has to make an appropriate choice of the
mass function as well as of the coefficient functions in the
ladder operators. We hope to address some of these issues
elsewhere.

\section*{Acknowledgements}
This work has been partially supported by Spanish Ministerio de
Educaci\'on y Ciencia (Project MTM2005-09183), Ministerio de
Asuntos Exteriores (AECI grant 0000147287  of A G),  and Junta de
Castilla y Le\'on (Excellence Project VA013C05). A~G~acknowledges
the authorities of City College, Kolkata, India for a study leave.


\end{document}